\def\be{\begin{equation}}
\def\ee{\end{equation}}
\def\te{\end{equation}}
\def\bea{\begin{eqnarray}}
\def\ba{\begin{eqnarray}}

\def\ta{\end{eqnarray}}
\def\tea{\end{eqnarray}}

\def\ben{\begin{enumerate}}
\def\een{\end{enumerate}}

\def\ha{{1\over 2}}

\def\n{\nu}

\def\bfr{{\bf r}}
\def\bfx{{\bf x}}
\def\bfy{{\bf y}}
\def\bfp{{\bf p}}
\def\bfL{{\bf L}}

\documentclass[12pt]{iopart}
  \expandafter\let\csname equation*\endcsname\relax
  \expandafter\let\csname endequation*\endcsname\relax
\usepackage{amsmath, amssymb}
\usepackage[hmargin=2.5cm,vmargin=2.5cm]{geometry}
\usepackage{graphicx}
\usepackage{enumerate}

\begin{document}

\title{Probing a Gravitational Cat State}
\author{C. Anastopoulos$^1$ and B. L. Hu$^2$}

\address{$^1$Department of Physics, University of Patras, 26500 Patras, Greece.}

\address{$^2$Maryland Center for Fundamental Physics and Joint Quantum Institute,\\ University of
Maryland, College Park, Maryland 20742-4111 U.S.A.}

\ead{anastop@physics.upatras.gr,blhu@umd.edu}

\date{April 10, 2015}

\begin{abstract}
We investigate the nature of  a gravitational two-state system (G2S) in the simplest setup in Newtonian gravity.  In a quantum description of  matter  a  single  motionless massive particle can in principle be in a superposition state of  two spatially-separated locations.  This superposition state in gravity, or gravitational cat state,  would lead to fluctuations in the Newtonian force exerted on a nearby test particle.   The central quantity of importance for this inquiry is the energy density correlation. This corresponds to the \textit{noise kernel in stochastic gravity theory},  evaluated in the weak field nonrelativistic limit.  In this limit quantum fluctuations of the stress energy tensor manifest as the  fluctuations of the Newtonian force. We describe the properties of such a G2S system and present two ways of measuring the  cat state for the Newtonian force, one by way of a classical probe, the other  a quantum harmonic oscillator.     Our findings  include:  (i)  mass density fluctuations persist even in single particle systems, and they are \textit{of the same order of magnitude as the mean};  (ii) a classical probe generically records a  \textit{non-Markovian fluctuating force}; (iii)  a quantum probe interacting with the G2S system may undergo Rabi oscillations in a \textit{strong coupling regime}. This simple prototypical gravitational quantum system could  provide a robust testing ground to \textit{compare predictions from alternative quantum theories}, since the results reported here are based on standard quantum mechanics and classical gravity.
\end{abstract}

\section{Introduction}

In this work we investigate the nature of a gravitational two-state system (G2S) in the simplest setup in Newtonian gravity.  In a quantum description of  matter  a  single  motionless massive particle can in principle be in a superposition state of  two spatially-separated locations.   This superposition state in gravity or gravitational cat state would lead to fluctuations in the Newtonian force exerted on a nearby test particle.    Though simple, this problem touches on three aspects, quantum, gravity and information.   We give a brief sketch of the backdrop of quantum and gravity,  then introduce the gravitational cat state and ask how to characterize and detect such a state in this setup.

\subsection{Backdrop of Quantum and Gravity}

Gravity and quantum are two cornerstones of modern physics which  for a good part of the last century are the source of awe and wonderment in their amazingly wide range of  applicability,  and at the same time the source of  persistent challenge in how to fuse them together  into a unified theory.  Finding ways to unify quantum (Q) and gravitation (G), each having stood the test of time with great successes in their respective realms, is the goal of  ``quantum gravity" programs (schematically denoted as  Q $\otimes$ G) , which has met with varying degrees of success (or deficiency, depending on one's view and values).   A more modest yet no less productive attempt is  to place these two theories closely together (schematically denoted as Q $\oplus$ G),  see what discrepancies  this union may reveal,  what difficulties we face in the interpretations and what new physical insights we may gain.  Considering that gravitation, as embodied in Einstein's theory of general relativity (GR), is a classical theory for  the large scale structure and dynamics of spacetime,  and quantum field theory (QFT) for  quantum matter down to the microscopic scale,  this task is far from  straightforward. The final analysis requires addressing the issues of quantum versus classical and micro versus macro  which  underlie almost all physical problems  in contemporary physics.

The theoretical effort of this vein can be traced back to the 60-70's in {\em quantum field theory in curved spacetime} (QFTCST) \cite{BirDav,Wald,ParTom} which began in the late 60's with cosmological particle creation studies  and epitomized in  Hawking's 1974  discovery of black hole radiance \cite{Hawk}. Focused efforts in seeking ways to regularize or renormalize the stress energy tensor of quantum fields made it possible to tackle the so-called `backreaction problem' 
in finding how quantum matter fields affect the dynamics of spacetime.  Solving the backreaction problem is at the core of {\em semiclassical gravity} theory (SCG) 
developed in the 80's based on the \textit{semiclassical Einstein equation }(SCE)\footnote{For references in these developments, see, e.g., the bibliographies in the monographs listed above and  references listed in a recent overview \cite{EmQM13} with relevance  to the issues addressed in this paper.}.

Following this vein, general relativists have probed the interplay between gravity and quantum largely from the angle of how quantum matter affects spacetime (Q $\rightarrow$ G). Asking the question in the other direction (G$\rightarrow$ Q), namely, how gravity could have an effect on quantum phenomena, has been going on for just as long (e.g., Ref. \cite{Karol})
mainly by quantum foundation theorists.  The foremost issue is why  macroscopic objects are found sharply localized in space (their wave functions ``collapsed" on definite locales) while those of microscopic objects extend over space. This contradiction is captured in the celebrated Cat of Schr\"odinger. One can very coarsely place these theories in three groups: The Ghirardi-Rimini-Weber - Pearle  models \cite{GRWP}
of  continuous spontaneous localization,  the Diosi-Penrose theories \cite{Diosi,Penrose}
invoking gravitational decoherence,  and the recent  trace dynamics theory of Adler \cite{AdlerCUP}
which attempts to provide a sub-stratum theory from which quantum mechanics emerges. A nice description of these theories can be found in a recent review by Bassi et al \cite{BassiRMP}.

These efforts spaning several decades are now called back to attention mainly due to  the fast development of experimental capabilities to put these theories to tests.  In fact, a new name ``gravitational quantum physics" (see, e.g., Ref.  \cite{NJPgqp} )  is coined by experimentalists  to capture the convergence of interest  from both the quantum foundation and the quantum + gravity communities.  The  interesting array of ongoing experiments using atomic-optical, molecular spectroscopy, nanomechanical and other schemes to test alternative quantum theories and/or the effects of gravity on quantum systems also bring in lively issues in the quantum-classical and micro-macro interfaces.  In this context,  issues related to quantum information such as quantum decoherence and entanglement arise. This last aspect is the target scope of our present paper.

\subsection{Quantum Superpositions involving Gravity}

We now add quantum information considerations to  quantum systems involving gravitational interaction.  A simple problem of foundational interest is the superposition of two macroscopically distinct states.
This simple yet basic state, which goes by the name of a ``cat state",  has been studied and tested out for some nongravitational systems \cite{Monroe}.  Experimental schemes for testing quantum superposition of mechanical objects (e.g., mirrors, cantilevers) by their interaction with light and electric devices  have also been  explored \cite{Marshall,NJPqom,Chen,CavityOM} in the frontline research areas of opto- and nano-electro-mechanics.

As illustration of a gravitational cat state,  consider the quantum description of a  stationary point mass $m$  localized around $\bf x = 0$ with spread $\sigma$, described by a Gaussian wave function with zero mean momentum.
 \begin{eqnarray}
\psi_0(\bfx) =  \frac{1}{(2 \pi \sigma^2)^{3/4}} {e^{ - \frac{{\bfx}^2}{4\sigma^2}} }. \label{Gausswfn}
\end{eqnarray}
The position ${\bfx}$ of the particle is a random variable described by the probability distribution $|\psi_0(\bfx)|^2$. However, according to Newton's law, a probability distribution for ${\bf x}$, defines a probability distribution for the Newtonian force acted on a particle of mass $m_0$ located at ${\bf R}$
\begin{eqnarray}
{\bf F} = - \frac{Gmm_0}{|\bf R - \bf x|^3}(\bf R - \bf x). \label{newtforce}
\end{eqnarray}
 For $|{\bf R}| >> \sigma$ the fluctuations of the Newtonian force are negligible,  which leads one to view it as a deterministic variable.


 Now consider a cat state, i.e., a superposition of two Gaussians, each  located at $\pm \ha {\bf L}$ and  with zero mean momentum,
 \begin{eqnarray}
\psi(\bfx) = \frac{1}{\sqrt{2}} \frac{1}{(2 \pi \sigma^2)^{3/4}} \left[e^{ - \frac{(\bfx - \bfL/2)^2}{4\sigma^2}} +     e^{ - \frac{(\bfx - \bfL/2)^2}{4\sigma^2}}         \right]  \label{catwfn}
\end{eqnarray}
If ${\bf L}$ is of the order of magnitude of ${\bf R}$, the fluctuations of the Newtonian force (\ref{newtforce}) are non-negligible. Since the force is a function of ${\bf x}$, and ${\bf x}$ is described by an operator in quantum mechanics, the Newtonian force should also be described as an operator. But then, so would be the gravitational potential.  In this sense, the cat state for the point mass has generated a cat state for the gravitational field.

What is the right theory to use for the description of  a   gravitational cat state? What characteristics do they possess and how they can be probed are the questions we ask in this paper.  These inquires are the starting point of any investigation into quantum informational issues of gravitational systems.

\subsection{What is the right theory to use?}

In answer to the first question above,   we may begin  searching for a theory which describes quantum matter existing / propagating  in a classical gravitational  field, or QFT in curved spacetime.  Further demanding self-consistency in the solutions of  the dynamical equations for both classical gravity and for quantum matter we reach the semiclassical gravity theory described by the semiclassical Einstein equation (SCE), (also known as the M\/oller-Rosenfeld equation \cite{MolRos} in the alternative quantum theories community):
\begin{equation}
G_{\mu\nu} = 8 \pi G <T_{\mu\nu}(\Phi)>  \label{SCEq}
\end{equation}
 where $T_{\mu\nu}(\Phi) $ is the stress energy tensor of the matter field, here represented by a scalar field $\Phi$, and $< >$ denotes taking the expectation value with respect to a certain quantum state, usually the vacuum state.

However,  we run into a problem immediately.  Consider a single particle. For a superposition of two states each localized  at $\bfx= \pm \ha \bfL$, the SCE eqn makes the prediction that the particle is most likely to be found at the center $\bfx = 0$,  rather than at  either side.   This intrinsic deficiency of SCG has been pointed out in e.g., \cite{Ford82,QG2}. Some authors \cite{PagGei,EppHan} use this as  an indication that gravity needs be quantized\footnote{Note the subtle yet important distinction between quantizing the weak perturbations of gravity (gravitons as spin 2 fields)  and quantum gravity, defined as theories for the microscopic structures of spacetime.  The former is on the same footing as phonons in regard to the latter as atoms \cite{E/QG}.  The former can exist at today's low energy while the latter pertains to Planck scale physics which we have limited facts to speculate on. }.  One needs to include considerations of the quantum correlations, or equivalently, quantum fluctuations of the mass density of the system to be able to capture the physics of the superposition state.

There is such a theory, known as stochastic gravity introduced in the 90's whose main challenge was to find a lawful place for the fluctuations of quantum fields as a part of the  total source driving the Einstein equation.  Adding  the expectation value of the stress energy bitensor, also known as the noise kernel,  to the expectation value of the stress energy tensor upgrades the SCEq to  the \textit{Einstein-Langevin equation} \cite{ELE}, which enables one to solve for the induced metric fluctuations (the poetic  `spacetime foam') in a self consistent manner.    Stochastic gravity
\cite{stograHuVer} has been developed and applied to strong field situations such as structure formation in the early universe \cite{RouVer08} and black hole fluctuation and backreaction issues \cite{HuRou07}.  Here,  in the opposite regime,  of weak field  and nonrelativistic limit, stochastic gravity based on GR + QFT is also  the theory suitable for addressing quantum information issues at laboratory accessible low energy ranges  \footnote{In terms of theoretical structure, stochastic gravity occupies the intermediate range as one progresses (`bottom-up') from semiclassical  to quantum gravity, in the sense that when  higher correlation functions of  quantum matter fields are added as source of the Einstein equation the quantum coherence of the gravity sector begins to appear and increase in degree with the order of the correlation hierarchy. See description in \cite{kinQG}. The SCG as a mean field approximation capturing only the lowest order is understandably inadequate in this regard. }.

For readers familiar with the essentials of stochastic gravity theory,  the main character in our exposition, the mass density correlator, is just the weak field nonrelativistic limit of the noise kernel.   However, our presentation below does not assume any familiarity of stochastic gravity. It should be  understandable with some  minimal knowledge  of nonrelativistic quantum fields and  classical gravity.

\subsection{Gravitational Cat State}

In non-relativistic gravity, a mass density distribution $\mu$  generates a gravitational potential $\phi$ via the Poisson equation
\begin{eqnarray}
\nabla^2 \phi = -4 \pi G \mu.
\end{eqnarray}
A fluctuating mass density implies a fluctuating potential and thus a fluctuating force on any test particle. Hence, a quantum state that corresponds to a superposition of states with different mass density distributions gives rise to a `superposition' state of gravitational potentials,  i.e., a gravitational cat state.

Gravitational cat states are not new. The reason they are not in our conscious state of mind,  much less the object of serious study, is likely because they have been viewed as undesirable, as targets of elimination, from the start.  This is the spirit of many alternative quantum theories \cite{BassiRMP}.

 Gravitational cat states are rarely mentioned in their proper setting, from quantum field theory in curved spacetime to quantum gravity, because these theories mostly aim at describing ultra-high energy processes near the Planck scale. Only in the field of quantum cosmology in the context of the ``wave function of the universe" \cite{HarHaw}  or  the ``universal wave function" (many-worlds interpretation) \cite{Everitt} is an analogous problem discussed, because the generic solution to the Wheeler-DeWitt equation is a cat state, i.e., a superposition of two distinct cosmological spacetimes---see, for example, Ref. \cite{QC2}.   Our aim here is to explore the nature and consequences of gravitational cat states in the low energy realm that is more amenable to laboratory experiments.

 Gravitational two state (G2S) systems  should play a prominent role in the new field called ``gravitational quantum physics" \cite{NJPgqp},  ushered in by leading experimentalists working on atomic-molecular, opto- and electro-mechanical schemes wanting to explore issues involving quantum, information and gravitation.

 \subsection{Gravitational Decoherence}

It is perhaps useful to recall  Penrose's influential argument against the existence of gravitational cat states \cite{Penrose}, that is employed in support for  gravitational decoherence. It proceeds as follows.

In quantum theory, time is an external parameter manifested in the evolution of the quantum states through Schr\"odinger's equation. In general relativity, time is part of the spacetime structure, which is dynamical and depends on the configuration of the matter degrees of freedom. This contradiction is manifested clearly when considering superposition of macroscopically distinct states. Each component of the superposition generates a different spacetime. Since there is no canonical way of relating time parameters in different spacetime manifolds, there is a fundamental ambiguity in the choice of the time parameter in the evolution of the quantum state. This ambiguity is manifested even at low energies, when the gravitational interaction can be effectively described by the Newtonian theory. It suggests a physical mechanism of gravity-induced decoherence for superpositions of macroscopically distinct states.

We agree that the ambiguity in the definition of the time coordinate in a cat state is real, as far as our present knowledge of fundamental physics is concerned. In fact, we  have rephrased this argument in terms of gauge independence for the symmetry of space and time reparameterizations in Ref. \cite{AnHu13}. However,  the conclusion that a cat state must decohere does not follow. As we have argued in Ref. \cite{AnHu08}, strong additional assumptions about the character of such  ambiguities or fluctuations are needed in order to infer that they are agents of decoherence.   Schr\"odinger's cat  may very well coexist peacefully with temporal ambiguities or fluctuations.

\subsection{The present study}

We study the stress energy tensor and its fluctuations at the level of a single non-relativistic particle. In the non-relativistic limit, the dominant contribution to the stress-energy tensor comes from the mass density $\mu(\bfr)$.  The corresponding quantum operator is proportional to the number density operator  that defined in terms of the particle creation and annihilation operators, or equivalently, the non-relativistic quantum fields. Thus, expectation values and correlations of the mass density can be defined as expectation values of products of $\hat{\mu}(\bfr,t)$.

\subsubsection{ Properties of fluctuations.} With the above characterizations, we establish two important properties of the mass density fluctuations.

\medskip

First, {\em mass density fluctuations are very strong in single-particle systems}. They are of the same order of magnitude as the mean mass density.
This effect, though somewhat unusal, is in line with the earlier findings of  Ford, Hu and their co-workers \cite{KuoFor94,PH97,PH00}, where the fluctuations of the energy density of quantum fields in Minkowski space, Casimir geometry, Einstein universe are found to be of the order of their mean values. This was one of the reasons why fluctuations need be included in the consideration of the influence of quantum matters on spacetime structure and dynamics, as highlighted in stochastic gravity theory.

This result is  in contrast to the nature of mass-density as defined  in classical hydrodynamics, where fluctuations are usually assumed negligible. However, the  suppression of fluctuations in hydrodynamics is statistical. It arises out of the large number of particles constituting the physical system and quantum fluctuations are usually unimportant in comparison to statistical fluctuations. However, when considering a single- or few-particle system statistical fluctuations on the number of particles in a given region, while important,  do not, in general, dominate over quantum meachanical fluctuations.

\medskip

 Second, there exists a relation between   the correlation functions of the mass density and  position histories of the associated quantum particle.   The correlation functions of the mass density correspond to values of the {\em decoherence functional} for position histories of the particle. The decoherence functional is a complex functional of pairs of histories that appears in the consistent/decoherent histories approach to quantum mechanics \cite{Omn1, Omn2, Gri, GeHa, hartlelo}.

\subsubsection{Probing a Gravitational  Two-State (G2S) System.}

We consider a particle prepared in a cat state associated with the two minima of a confining potential and we calculate the corresponding gravitational force.  Two cases are considered. The first corresponds to a classical probe that records continuously the gravitational force. We compute the full probability distribution for the force measurements, and we show that they satisfy an exponential law and are distinctly non-Markovian. The second corresponds to a quantum probe, an harmonic oscillator acted upon by the Newtonian  force. We show that the system is described by the Jaynes-Cummings Hamiltonian, and we determine the probe's evolution in the relevant regime. We find that tunneling between the two minima of the potential leads to Rabi oscillations for the  probe.

\subsection{Organization of this paper}

We will first describe in Sec. 2 the  stress energy correlations in nonrelativistic field theory, and we show their relation to the decoherence functional and to the Wigner function that describe the  time evolution of a single particle. In Sec. 3 we consider a quantum system characterized by a confining potential with two local minima such as a double well potential. In Sec. 4 we present two ways to probe the properties of such a gravitational two level system.  In Sec. 5 we summarize the important features brought out in our analysis of this prototypical gravitational quantum system.


\section{Stress-energy correlations}

\subsection{Nonrelativistic N Particle System}

Consider a scalar quantum field $\hat{\phi}(\bfr)$ and its conjugate momentum $\hat{\pi}(\bfr)$ expressed in terms of the creation and annihilation operators $\hat{a}_{\bf k}$ and $\hat{a}^{\dagger}_{\bf k}$
\begin{eqnarray}
\hat{\phi}(\bfr) = \int \frac{d^3k}{(2\pi)^3 \sqrt{2 \omega_{\bf k}}} \left[ \hat{a}_{\bf k} e^{i {\bf k}\cdot \bfr} + \hat{a}^{\dagger}_{\bf k} e^{- i {\bf k}\cdot \bfr}\right] \label{fieldq1} \\
\hat{\pi}(\bfr) = i \int \frac{d^3k}{(2\pi)^3 } \sqrt{\frac{ \omega_{\bf k}}{2}}\left[ - \hat{a}_{\bf k} e^{i {\bf k}\cdot \bfr} + \hat{a}^{\dagger}_{\bf k} e^{- i {\bf k}\cdot \bfr}\right]. \label{fieldq2}
\end{eqnarray}
For a free field, the Hamiltonian operator is
\begin{eqnarray}
\hat{\rm H} = \int \frac{d^3k}{(2\pi)^3 }  \omega_{\bf k} \hat{a}^{\dagger}_{\bf k} \hat{a}_{\bf k},
\end{eqnarray}
where $\omega_{\bf k} = \sqrt{{\bf k}^2 + m^2}$.

In the non-relativistic regime we define the fields
\begin{eqnarray}
\hat{\psi}(\bfr) = \int   \frac{d^3k}{(2\pi)^3 } \hat{a}_{\bf k} e^{i {\bf k}\cdot \bfr} , \hspace{1cm}
\hat{\psi}^{\dagger}(\bfr) = \int \frac{d^3k}{(2\pi)^3 } \hat{a}^{\dagger}_{\bf k} e^{- i {\bf k}\cdot \bfr}, \label{psii}
\end{eqnarray}
and to leading order in $|{\bf k}|/m$,
\begin{eqnarray}
\hat{\phi}(\bfr) = \frac{1}{\sqrt{2m}} \left[ \hat{\psi}(\bfr) + \hat{\psi}^{\dagger}(\bfr) \right],   \hspace{1cm}
\hat{\pi}(\bfr) = - i \sqrt{\frac{m}{2}} \left[ \hat{\psi}(\bfr) - \hat{\psi}^{\dagger}(\bfr) \right]. \label{newt}
\end{eqnarray}
The Hamiltonian then becomes
\begin{eqnarray}
\hat{\rm H} = m \int d \bfr \hat{\psi}^{\dagger}(\bfr) \hat{\psi}(\bfr)  - \frac{1}{2m}\int d \bfr \hat{\psi}^{\dagger}(\bfr) \nabla^2 \hat{\psi}(\bfr). \label{hamnr}
\end{eqnarray}
We will denote the second term in Eq.(\ref{hamnr}) as $\hat{H}_0$ because it corresponds to the Hamiltonian for $N$ non-relativistic particles.  The first term in Eq.(\ref{hamnr}) corresponds to $N m$, for an $N$-particle state. Hence, the number operator $\hat{N}$ is
\begin{eqnarray}
\hat{N} = \int d \bfr  \hat{\psi}^{\dagger}(\bfr) \hat{\psi}(\bfr) \label{numop}
 \end{eqnarray}
This suggests that $ m \hat{\psi}^{\dagger}(\bfr) \hat{\psi}(\bfr)$ can be identified as the mass-density operator $\hat{\mu}(\bfr)$.

We include the effect of a confining potential $V({\bfr})$ , by modifying the  field Hamiltonian
\begin{eqnarray}
\hat{\rm H} = m \int d \bfr \hat{\psi}^{\dagger}(\bfr) \hat{\psi}(\bfr)  + \int d \bfr \hat{\psi}^{\dagger}(\bfr) \left[- \frac{1}{2m} \nabla^2 + V(\bf r)\right] \hat{\psi}(\bfr).
\end{eqnarray}

\subsection{Mass-density correlations, noise kernel}

  In the non relativistic limit, the dominant component of the stress tensor $T_{\mu \nu}$ is the energy density, which is dominated by the mass density, namely
  \begin{eqnarray}
  T_{\mu \nu}(\bfr, t) = \delta^0_{\mu} \delta^0_{\nu} \mu(\bfr, t)
  \end{eqnarray}
Thus, it suffices to calculate the correlation functions of the Heisenberg-picture operator
\begin{eqnarray}
\hat{\mu}(\bfr, t) = e^{i\hat{H}t} \hat{\mu}(\bfr) e^{-i\hat{H}t}.
\end{eqnarray}
We assume an one-particle state
  \begin{eqnarray}
  |\phi\rangle = \int d \bfr \phi(\bfr) \hat{\psi}^{\dagger}(\bfr)|0\rangle ,
  \end{eqnarray}
where $\phi(\bfr)$ is the one-particle wave-function.

We find
  \begin{eqnarray}
  \langle \mu(\bfr,t)\rangle = \langle \phi| \hat{\mu}(\bfr, t)|\phi\rangle = m \phi^*(\bfr, t) \phi(\bfr,t)\\
   \langle \mu(\bfr,t) \mu(\bfr',t')\rangle = \phi^*(\bfr, t) \phi(\bfr',t') G(\bfr, t;\bfr',t') \label{corr2}
  \end{eqnarray}
where $\phi(\bfr,t)$ is the time-evolved single particle wave function and $G(\bfr, t;\bfr',t')$ is the one-particle propagator,
 \begin{eqnarray}
 G(\bfr, t; \bfr', t') = \langle \bfr'|e^{-i\hat{H}(t' -t)} |\bfr\rangle.
 \end{eqnarray}
For a free particle,
 \begin{eqnarray}
 G(\bfr, t;\bfr',t') = \left(\frac{m}{2\pi i t}\right)^{3/2} \exp \left[ \frac{im(\bfr - \bfr')^2}{2(t'-t)}\right]
 \end{eqnarray}

 We note that the two-point correlation function is complex valued and cannot define a stochastic process.   However, the real part,
 \begin{eqnarray}
 \xi(\bfr, t;\bfr',t') = Re    \langle \mu(\bfr,t) \mu(\bfr',t')\rangle,
 \end{eqnarray}
known as the noise kernel, corresponds in some cases  to the two-point correlation function of a stochastic process.

Of importance is also the connected two-point correlation function for the mass densities
\begin{eqnarray}
\eta (\bfr,t; \bfr',t') =  \langle \mu(\bfr,t) \mu(\bfr',t')\rangle  - \langle \mu(\bfr,t)\rangle \langle \mu(\bfr', t')\rangle.
\end{eqnarray}

\subsection{Relation to the decoherence functional}

Let us ponder for a moment the meaning of the mathematical expressions above. Note first that the expectation values  (\ref{corr2}) do not correspond to the correlation functions of a physical process, because in realistic systems the mass density is not defined at a sharp spacetime point but smeared  over a finite spacetime region. In actual experiments, the particles under consideration (atoms) have a finite size \textsl{d} and it is meaningless to talk about mass densities at scales smaller than \textsl{d}, unless one has a detailed knowledge of the particle's internal state.

For this reason, rather than the exact mass density function, we consider a smeared mass density function (for more about applying smearing function for the evaluation of distributions, see, e.g., \cite{PH00,HuRou07})
\begin{eqnarray}
\hat{\mu}_s (\bfr, t) = \int d \bfr' f(\bfr - \bfr') \hat{\mu}(\bfr', t), \label{smearedmu}
\end{eqnarray}
for some smearing function  $f(\bfr)$ of dimension $[\mbox{length}]^{-3}$, centered around $\bfr = 0$. The smearing scale $\ell$ is defined by the condition $\ell^3 = 1/f(0)$.

We define the positive operator
\begin{eqnarray}
\hat{P}_{\bfr} = \int d \bfr' g(\bfr- \bfr') |\bfr'\rangle \langle \bfr'|,
\end{eqnarray}
where $g(\bfr) := f(\bfr)/f(0)$.  The operator $\hat{P}_{\bfr} $ represents a sampling of position with a width $s_x$ around $\bfr$.

If the sampling function $g$ is a characteristic function of some set (i.e., if $g^2 = g$), then  $\hat{P}_{\bfr}$ is a projection operator. Here, we will consider Gaussian functions of the form
$g(\bfr) = e^{- \frac{\bfr^2}{2 s_x^2}}$.
in which case  $\hat{P}_{\bfr}$ is an approximate projector. The corresponding  smearing function is
\begin{eqnarray}
f(\bfr) = \frac{1}{(2 \pi)^{3/2}s_x^3} e^{- \frac{\bfr^2}{2 s_x^2}} \label{smearing}
\end{eqnarray}
and corresponds to $\ell = \sqrt{2\pi}s_x$.

The correlation functions of the mass density become
\begin{eqnarray}
\langle \mu_s(\bfr, t) \rangle = \frac{m}{\ell^3} \langle \phi|\hat{P}_{\bfr, t} |\phi\rangle \\
\langle \mu_s(\bfr, t) \mu_s(\bfr',t') \rangle  = \frac{m^2}{\ell^6} \langle \phi| \hat{P}_{\bfr t} \hat{P}_{\bfr' t'}                  |\phi\rangle, \label{smearedcorr}
\end{eqnarray}
where $\hat{P}_{\bfr t} = e^{i \hat{H} t} \hat{P}_{\bfr} e^{-i\hat{H}t}$ is the Heisenberg-picture evolution of $\hat{P}_{\bfr}$.

The expectation value of the smeared mass density is proportional to the probability of a position measurement at time $t$.  The two point correlation function is proportional to the decoherence functional
\begin{eqnarray}
{\cal D}(\bfr, t; \bfr', t') := \langle \phi| \hat{P}_{\bfr t} \hat{P}_{\bfr' t'}                  |\phi\rangle   \label{dfun}
 \end{eqnarray}
 for  a pair of histories  one corresponding to  a position record $\bfr$ at time $t$ and the other to a position record $\bfr'$ at time $t'$. As explained by the decoherent histories approach to quantum mechanics, the vanishing of the off-diagonal elements of the decoherence functional in an exhaustive and exclusive set of histories is a necessary condition for assigning probabilities to this set of histories. The assignment of probabilities implies that the time evolution of the system can be expressed in terms of  classical equations of motion, possibly including a stochastic component.

  Typically  an approximate projector remains   close to a true projector under time evolution, so
  $\hat{P}_{\bfr t}^2 \simeq  \hat{P}_{\bfr t}$. Of course, the equality is exact if $\hat{P}_{\bfr}$ is a sharp projector. Then,  $\langle \mu_s(\bfr, t) \mu_s(\bfr,t) \rangle = \frac{m}{\ell^3} \langle \mu_s(\bfr, t) \rangle$, and the connected correlation function becomes
\begin{eqnarray}
\eta (\bfr,t; \bfr,t)  = \frac{m}{\ell^3}
\langle \mu_s(\bfr, t) \rangle - \langle \mu_s(\bfr, t) \rangle^2 \geq 0.
\end{eqnarray}
For $t = t'$, ${\cal D}(\bfr, t; \bfr', t) \sim \delta (\bfr - \bfr')$; thus, the correlation function $\langle \mu_s(\bfr, t) \mu_s(\bfr',t) \rangle$ vanishes at equal times, unless $\bfr = \bfr'$ within an accuracy of order $\ell$.

The values of  the decoherence functional ${\cal D}(\bfr, t; \bfr', t')$ for general arguments  can differ significantly  from zero only if the vectors  $\hat{P}_{\bfr t}|\phi\rangle $ and  $\hat{P}_{\bfr' t'}|\phi\rangle $ overlap substantially. Consider as an example the case of a free particle and assume that
$|\phi\rangle$ is well localized in position around some point $\bfr_0$. The vector $\hat{P}_{\bfr t}|\phi\rangle $ is  supported only on the momentum components $\bf p$ of $|\phi\rangle$ such that $\bfp = m (\bfr - \bfr_0)/t$, and similarly $\hat{P}_{\bfr' t'}|\phi\rangle $ is supported only on the momentum components $\bfp$ of $|\phi\rangle$ such that $\bfp = m (\bfr' - \bfr_0)/t'$. The two vectors overlap if $(\bfr - \bfr_0)/t \simeq (\bfr' - \bfr_0)/t'$. If this condition is satisfied, the values of the decoherence functional  are determined by the
  time-of-flight momentum
\begin{eqnarray}
\bfp_{TF} = m \frac{\bfr' - \bfr'}{t-t'}. \label{tofm}
\end{eqnarray}
as
\begin{eqnarray}
{\cal D}(\bfr, t; \bfr', t') \simeq |\langle \phi|  \bfp_{TF}\rangle|^2 \delta [(\bfr - \bfr_0)/t - (\bfr' - \bfr_0)/t']. \label{decf}
\end{eqnarray}
The associated connected correlation function is   very different from a delta function on $t$ and $\bfr$; hence, fluctuations exhibit highly non-Markovian behavior.

 We note that Eq. (\ref{decf}) is relevant for well localized initial states, and does not apply, for instance, to a cat state.


\subsection{Wigner-Weyl representation of correlation functions}

Next, we proceed to evaluate the smeared correlation functions (\ref{smearedcorr}), using  the Wigner-Weyl representation, in which an operator $\hat{A}$ on the Hilbert space of a particle is represented by a function $F_{\hat{A}}$ on the associated state space.
\begin{eqnarray}
F_{\hat{A}}({\bf x}, {\bf p}) =  \int d{\bf y} \langle \bfx - \frac{\bfy}{2}| \hat{A}| \bfx + \frac{\bfy}{2}\rangle e^{i\bfp \cdot \bfy} . \label{weyltrans}
\end{eqnarray}
We denote the Wigner-Weyl transform of $\hat{P}_{\bfr,t}$ as $F_{\bfr,t}$. For a free particle, $F_{\bfr,t}$  is
\begin{eqnarray}
F_{\bfr,t} (\bfx, \bfp) =   g(\bfr - \bfx - \frac{\bfp t}{m}).  \label{frt}
\end{eqnarray}
The  Wigner-Weyl transform $F_{\bfr,t;\bfr',t'}$  of the product $\hat{P}_{\bfr t} \hat{P}_{\bfr' t'}$  that appears in Eq. (\ref{smearedcorr}) is approximated by

\begin{eqnarray}
F_{\bfr,t;\bfr',t'} (\bfx, \bfp) \simeq F_{\bfr,t} (\bfx, \bfp) F_{\bfr',t'} (\bfx, \bfp).  \label{quasiclassical}
\end{eqnarray}
 The approximation (\ref{quasiclassical}) is meaningful  in a quasi-classical regime where $\hat{x}$ and $\hat{p}$ correspond to coarse-grained observables, like, for example, the position and momentum of the center of mass of a composite particle.

With a smeared mass density function (\ref{smearedmu}) this yields,
\begin{eqnarray}
F_{\bfr,t;\bfr',t'} (\bfx, \bfp)  =    \exp \left[ - \frac{1}{s_x^2}\left( \bfx - \frac{\bfr + \bfr'}{2} + \frac{\bfp(t+t')}{2m}\right)^2
\right. \nonumber \\
\left. - \frac{(t-t')^2}{4m^2 s_x^2}\left( \bfp - m \frac{\bfr - \bfr'}{t-t'}\right)^2  \right]
\label{WignerF}
\end{eqnarray}
Substituting into  Eqs. (\ref{smearedcorr}), we obtain
\begin{eqnarray}
\langle \mu_s(\bfr, t) \rangle = \frac{m}{\ell^3} \int \frac{d\bfx d \bfp}{(2\pi)^3} W_0(\bfx,\bfp) F_{\bfr,t} (\bfx, \bfp) \\
\langle \mu_s(\bfr, t) \mu_s(\bfr',t') \rangle  =  \frac{m^2}{\ell^6} \int \frac{d\bfx d \bfp}{(2\pi)^3} W_0(\bfx,\bfp) F_{\bfr,t;\bfr',t'} (\bfx, \bfp) .
\end{eqnarray}
where $W_0$ is the Wigner function of the initial state.

 For scales of observation much larger than $s_x$, we   substitute the Gaussians in Eq. (\ref{frt}) by delta functions
\begin{eqnarray}
F_{\bfr,t} (\bfx, \bfp) &\simeq&  \ell^3  \delta(\bfr - \bfx - \frac{\bfp t}{m}) \\
F_{\bfr,t;\bfr',t'} (\bfx, \bfp) &\simeq&   \ell^6 \delta \left( \bfx - \frac{\bfr + \bfr'}{2} + \frac{\bfp(t+t')}{2m}\right) \delta(\bfr - \bfr' - \frac{\bfp}{m}(t-t')).
\end{eqnarray}
whence, with the use of (\ref{WignerF}),
\begin{eqnarray}
\langle \mu_s(\bfr, t) \rangle &=&   \frac{m}{(2\pi)^3} \int d \bfp W_0(\bfr - \frac{\bfp t}{m}, \bfp) \\
\langle \mu_s(\bfr, t) \mu_s(\bfr',t') \rangle &=&  \frac{m^5}{(2\pi)^3(t-t')^3} W_0(\frac{\bfr +\bfr'}{2} -  \frac{(\bfr - \bfr') (t +t')}{2(t-t')},  m \frac{\bfr - \bfr'}{t-t'}    ) \label{corwign}
\end{eqnarray}
The two-point correlation function is, therefore, straightforwardly calculated for any initial state by computing its Wigner function and setting $\bfx = \frac{\bfr +\bfr'}{2} -  \frac{(\bfr - \bfr') (t +t')}{2(t-t')}$ and $\bfp = m \frac{\bfr - \bfr'}{t-t'} $.

\subsubsection*{Special cases}

Next,    we consider an initial wave-function $\psi_0$ with vanishing mean momentum. We assume that the momentum spread is so small, that at  a macroscopic level of observation, the $p$-dependence of the Wigner function is essentially  that of a delta-function, $W_0(\bfr,\bfp) = (2\pi)^3 \delta(\bfp)|\psi_0(\bfr)|^2$. Then, we obtain
\begin{eqnarray}
\langle \mu_s(\bfr, t) \rangle =   m  |\psi_0(\bfr)|^2.  \label{mean1}\\
\langle \mu_s(\bfr, t) \mu_s(\bfr',t') \rangle = m^2 |\psi_0\left(\bfr \right)|^2 \delta^3(\bfr - \bfr').\label{2pt}
\end{eqnarray}
Both the expectation value of  and the two-point correlation function of $\hat{\mu}_s$ are time-independent for this class of initial states that includes both the Gaussian state (\ref{Gausswfn}) and the cat state (\ref{catwfn}). Both mimic stationary states within this approximation, because of the vanishing mean momentum. We also note that the delta function in Eq. (\ref{2pt}) is meaningful for variation of the arguments at scales much larger than $\ell$. For $\bfr \simeq \bfr'$, it is appropriate to substitute the delta function by
the smeared delta function $f(\bfr)$ as in Eq. (\ref{smearing}).

The expectation value $\langle \mu_s(\bfr, t) \rangle$ is equivalent to a classical mass density $\rho(\bfr) = m  |\psi_0(\bfr)|^2$. However, the two-point correlation function does not correspond to fluctuations of the classical mass-density $\rho(\bfr)$, because the latter would be proportional to $|\psi_0\left(\bfr \right)|^4$ rather than to $|\psi_0\left(\bfr \right)|^2$. Thus, the correlation function (\ref{2pt}) reveals a characteristically quantum property of the density fluctuations, with no classical analogue.

We estimate the relative size of the fluctuations by evaluating the ratio
\begin{eqnarray}
C(\bfr) = \left|   \frac{\eta(\bfr, t; \bfr, t)}{\langle \mu_s(\bfr, t) \rangle\langle \mu_s(\bfr, t) \rangle}             \right| = \left| \frac{1}{\ell^3 |\psi_0(\bfr)|^3} - 1           \right|.
 \end{eqnarray}
$C(\bfr)$ is not restricted to values much smaller than unity, thus, the correlations are of the same order of magnitude of the mean value, or even larger.

\subsection{Correlation functions for a stochastic process}

The quantum correlation functions of the mass-density are, in general, complex-valued, and they do not correspond to the correlation functions of a stochastic process. Such a correspondence would be highly desirable, because when considering the  measurement of such correlations, it is much more convenient to treat them as subject to the rules of classical probability theory

In general, the quantum correlation functions define a stochastic process only if specific decoherence conditions are satisfied \cite{Ana01}. In this case, it can be proven that, for Gaussian processes,  the real part of the quantum two-point correlation function (the noise kernel) is the two-point correlation function of a stochastic process. This implies that the observable is essentially classical and  its evolution can be modeled using  classical probability theory---see Refs. \cite{hartlelo, DoHa92, Ana01}. This is the case of the two-point function (\ref{2pt}), albeit in a rather trivial sense, because  the correlation function is time-independent.

The correspondence of the correlation functions to the decoherence functional suggests of a different way to define statistical correlation functions associated with the mass density. We define
\begin{eqnarray}
\langle \mu(\bfr_n,t_n) \ldots \mu(\bfr_2,t_2) \mu(\bfr_1,t_1) \rangle_{st} = \left(\frac{m}{\ell^3}\right)^n P_n(\bfr_1, t_1; \bfr_2, t_2; \ldots; \bfr_n t_n), \label{corrn}
\end{eqnarray}
where
\begin{eqnarray}
  P_n(\bfr_1, t_1; \bfr_2, t_2; \ldots; \bfr_n t_n) = Tr \left[\hat{P}_{\bfr_nt_n}\ldots \hat{P}_{\bfr_2t_2} \hat{P}_{\bfr_1t_1}\hat{\rho}_0 \hat{P}_{\bfr_1t_1} \hat{P}_{\bfr_2t_2} \ldots \hat{P}_{\bfr_nt_n}\right] \label{ntimepr}
\end{eqnarray}
is the probability that $n$ position measurements at times $t_i$, $i = 1, 2, \ldots, n$ yield values $\bfr_i$. The time-instants are ordered: $t_1 < t_2< \ldots < t_n$. Eq. (\ref{ntimepr}) applies if the operators  $\hat{P}_{\bfr t}$ are sharp projectors. If $\hat{P}_{\bfr t}$ are approximate projectors, we substitute each  $\hat{P}_{\bfr_i t_i}$ in Eq. (\ref{ntimepr}) with $\sqrt{\hat{P}_{\bfr_i t_i}}$.

The probabilities  (\ref{ntimepr})  are well defined and they may be employed in order to calculate the response of macroscopic detectors to the quantum system. However, they do not define a stochastic process, because they do not satisfy the Kolmogorov additivity condition that is essential for the definition of a stochastic process,
\begin{eqnarray}
  P_{n-1}(\bfr_2, t_2; \ldots; \bfr_n t_n) = \int d\bfr_1   P_n(\bfr_1, t_1; \bfr_2, t_2; \ldots; \bfr_n t_n).
\end{eqnarray}
Furthermore, the  probabilities (\ref{ntimepr}) are strongly dependent on the coarse-graining scale $\ell$ \cite{Ana04, Ana05}. In general, $\ell$ is a scale that is determined not only by the characteristics of the measured system (e.g. the size of an atom) but also from the method of sampling and precision of measurement of its position. This implies that the probabilities  (\ref{ntimepr}), and, consequently, the statistical correlation functions (\ref{corrn}) depend on the specific set-up  by which the mass density is being measured. In particular, they depend both on the sampling length $\ell$, but also on `how often' the system is being sampled, i.e., the number $n$ of measurements as in (\ref{ntimepr}) in a given time-interval. Hence, unless we consider systems in which the mass density behaves classically (as in (\ref{2pt})),
different measurement schemes will lead to different predictions for the fluctuations of the mass density, and, consequently, of the associated Newtonian force.

\section{A gravitational two-state system}

In this section, we consider  a quantum system characterized by a confining potential with two local minima located at $\bfr = \pm \frac{1}{2} \bfL$. We label the minima as $+$ and $-$. At the macroscopic level, we can only distinguish whether the particle lies in the $+$ region or in the  $-$region.  We take  the direction of $\bfL$ to correspond to the $x$ axis. Then, the projection operators onto the $\pm$ regions are given respectively by
\begin{eqnarray}
\hat{P}_+ = \int_0^{\infty} dx \int_{\-\infty}^{\infty}  dy \int_{\-\infty}^{\infty} dz |x,y,z\rangle\langle x,y,z| \nonumber \\
\hat{P}_+ = \int_{-\infty}^0 dx \int_{\-\infty}^{\infty}  dy \int_{\-\infty}^{\infty} dz |x,y,z\rangle\langle x,y,z|. \label{p+-}
\end{eqnarray}
 By this definition,
 $\hat{P}_+ + \hat{P}_- = \hat{1}$. We will employ indices $a, b, c$ and so on, that take two values $\pm $  in order to denote the two minima.

 We consider an initial state
\begin{eqnarray}
|\psi \rangle = c_+ |+\rangle + c_- |-\rangle.
\end{eqnarray}
 where $|+ \rangle$ and $|-\rangle$ are  states localized around the minima $+$ and $-$ respectively, i.e., they satisfy
 \begin{eqnarray}
 \hat{P}_a |a\rangle = \delta_{ab} |a\rangle,  \hspace{0.2cm} a,b = \pm. \label{pab}
 \end{eqnarray}
  Thus, the system can be expressed as a  quantum two state system (or qubit), where
 \begin{eqnarray}
 |+\rangle = \left(\begin{array}{c}1\\0\end{array}\right) \hspace{1cm}   |-\rangle = \left(\begin{array}{c}0 \\ 1\end{array}\right)
 \end{eqnarray}
In this approximation, it makes no difference whether we consider the projectors $\hat{P}_\pm$ or any other projector that satisfies Eq. (\ref{pab}). The smearing scale $\ell$ is identified as the width of the finest projectors that satisfy Eq. (\ref{pab}). Hence, it is of the order of magnitude of the width of the localization region around the potential minima.

\subsection{Trivial dynamics}
First, we assume that the tunneling probability between the two minima is so low, as to be negligible. This implies that $\hat{U}_t^{\dagger} \hat{P}_a \hat{U}_t = \hat{P}_a$.
\begin{eqnarray}
\langle \mu_s(a, t) \rangle = |c_a|^2  \frac{m}{\ell^3}   \\
\langle \mu_s(a, t) \mu_s(b,t') \rangle  = \frac{m^2}{\ell^6} |c_a|^2 \delta_{ab}, \label{smearedcorr2}
\end{eqnarray}
for all $a, b = \pm$. Thus, the correlation functions are time-independent. A measurement at any moment of time (through the exerted gravitational force) will find the system in one of the minima with probability given by $|c_a|^2$. The $\delta_{ab}$ in the two-point correlation function signifies that the measurement collapses the wave-function completely and any subsequent measurement will find the system in the original minimum.

\subsection{Tunneling dynamics}

We consider a two-level system, in which tunneling effect is an important contribution to its dynamics.  A double-well system is typically characterized by  a time-scale $\epsilon$ at which the transition amplitudes stabilize. Hence, we  can describe the system's  evolution in terms of successive time-steps.  For a two-state system $|\pm\rangle $  the  evolution operator $\hat{u}$ for each time-step $\epsilon$ is
\begin{eqnarray}
\hat{u} = \left( \begin{array}{cc} \cos \theta &\sin \theta e^{i \chi}\\-\sin \theta e^{-i \chi} & \cos\theta  \end{array} \right),
\end{eqnarray}
for some angles $\theta > 0$ and $\chi$, where $\theta << 1$.

For $t >> \epsilon$, we take the  continuous-time limit by computing $\hat{u}^n$ and setting $n = t/\epsilon$. We obtain an one parameter family of unitaries
\begin{eqnarray}
\hat{U}_t= \left( \begin{array}{cc} \cos \frac{\nu t}{2}&\sin \frac{\nu t}{2} e^{i \chi}\\-\sin \frac{\nu t}{2} e^{-i \chi}&\cos \frac{\nu t}{2}  \end{array} \right),
\end{eqnarray}
for $\nu = 2\theta/\epsilon$, which correspond to a Hamiltonian
\begin{eqnarray}
\hat{H} = \nu (\cos\chi \hat{\sigma}_1 + \sin \chi \hat{\sigma}_2). \label{hamqub}
\end{eqnarray}

Then, the evolved projectors $\hat{P}_{\pm t} = \hat{U}^{\dagger}_t \hat{P}_{\pm} \hat{U}_t$ are
\begin{eqnarray}
\hat{P}_{\pm t} = \frac{1}{2}\left(\hat{1} + a \hat{S}_t\right), \\
\hat{S}_t = \left( \begin{array}{cc} \cos \nu t & \sin \nu t e^{i\chi} \\ \sin  \nu t e^{-i\chi} & -\cos \nu t\end{array} \right),
\end{eqnarray}
where $a =\pm 1$.

 The quantum correlation functions are
\begin{eqnarray}
\langle \mu(a, t) \rangle = \frac{m}{2\ell^3}[1 + a (\delta \cos \nu t + \beta \sin \nu t)] \\
\langle \mu(a_2, t_2) \mu(a_1, t_1) \rangle  = \frac{m^2}{4\ell^6}  \left[ 1 + \delta (a_1 \cos \nu t_1 + a_2 \cos \nu t_2 + a_1a_2 \cos\nu(t_2-t_1)) \right. \nonumber \\
\left. + \beta (a_1 \sin \nu t_1 + a_2 \sin \nu t_2) - i \gamma a_1 a_2 \sin \nu(t_2-t_1) \right].
\end{eqnarray}
where $a, a_1, a_2 = \pm$. The real numbers $\delta, \gamma$ and $\beta$ describe the initial state:   $\delta = |c_+|^2 - |c_-|^2$ is the asymmetry between the two minima, and $\beta + i \gamma = 2c_+^*c_-e^{i\chi}$.

We also calculate the statistical correlation function
\begin{eqnarray}
\langle \mu(a_2, t_2) \mu(a_1, t_1) \rangle_{st} = \frac{m^2}{4\ell^6} \left[ 1 + a_1(\delta \cos\nu t_1 + \beta \sin \nu t_1) \right. \nonumber \\
\left. + a_2 \cos \nu(t_2-t_1)(\cos \nu t_1 - \beta \sin \nu t_1) + a_1 a_2 \delta  \cos\nu(t_2-t_1) \right].
\end{eqnarray}

Again,  we see that the fluctuations of the mass density are of the same order of magnitude as the expectation values.

\section{Probing a gravitational two-state (G2S) system}

Next, we present two different ways that mass density fluctuations of the two-level system described in Sec. 3 can be measured. In the first case, the probe, or the measuring apparatus,  is a classical test-mass  that records the Newtonian force continuously. In the second case,  the probe is a quantum harmonic oscillator  coupled to the gravitational two-state system described above. We consider the gravitationally induced effects to the oscillator's evolution.

\subsection{Classical test- mass probe: Fluctuating force}

Consider a test  mass $m_0$ located near the confining potential.
We choose the coordinate system so that the two  minima of the potential lie at $(\pm\frac{L}{2}, 0, 0)$ and the test mass probe at $(0,y,0)$, as in Fig. 1.

\begin{figure}[tbp]
\includegraphics[height=8cm]{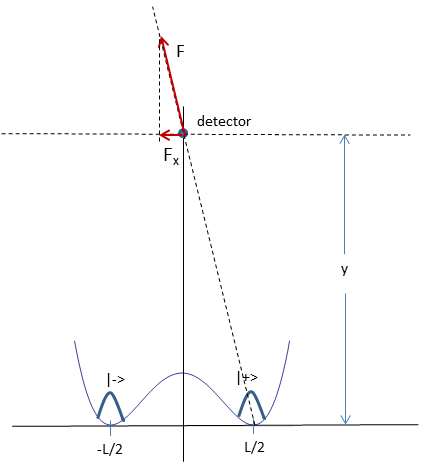} \caption{ \small  Force on a probe / detector exerted by a massive particle in a gravitational cat state between locations $\pm L/2$ .}
\end{figure}

We evaluate the $x$ component of the Newtonian force exerted by the two-level system on the test particle. Again denote $a = \pm$. If the system lies on the  minimum of the potential at $x = aL/2$, the force $F_x(a)$ exerted on the test particle in the x direction is
 \begin{eqnarray}
 F_x(a) = -\frac{Gm m_0L}{2(y^2+L^2/4)^{3/2}} a = - f_0 a \label{fxa}
 \end{eqnarray}
Assuming that the test mass is not allowed to move,  the force $F_x$ takes only two values $f_0$ and $-f_0$. These values are correlated with the projectors $\hat{P}_a$, Eq. (\ref{p+-}). Thus $F_x$ corresponds to a self-adjoint operator
\begin{eqnarray}
\hat{F} = -f_0 \hat{P}_+ + f_0 \hat{P}_- = - f_0 \hat{\sigma}_3, \label{fobs}
\end{eqnarray}
on the 2-state system's Hilbert space. Thus, the gravitational force  behaves as a quantum variable, its probabilities and correlations determined by quantum mechanics.

  Since Newton's law of gravity is instantaneous, the test mass will be acted upon by the gravitational force at all times. Hence, the measurement output will be a time series of recorded force values. Thus, this set-up corresponds to a {\em continuous measurement} \cite{contmeas} of the G2S system.

 In Appendix A, we calculate the correlation functions of the force, obtained from the quantum probabilities for $N$ successive measurements
 \begin{eqnarray}
  P(a_1, t_1; a_2, t_2; \ldots; a_N t_n) = Tr \left[\hat{P}_{a_Nt_N}\ldots \hat{P}_{a_2t_2} \hat{P}_{a_1t_1}\hat{\rho}_0 \hat{P}_{a_1t_1} \hat{P}_{a_2t_2} \ldots \hat{P}_{a_nt_n}\right] \label{ntimepr2}
\end{eqnarray}
  Without loss of generality, we consider an initial $|+\rangle$ state. We find
 \begin{eqnarray}
\langle F(t)\rangle = - f_0 e^{-\Gamma t} \label{corrcon1}
\\
\langle F(t') F(t)\rangle = f_0^2 e^{- \Gamma |t'-t|}. \label{corrcon2}
\end{eqnarray}
 The decay constant $\Gamma$ is defined as
 \begin{eqnarray}
  \Gamma = \frac{\nu^2 \tau}{2},
 \end{eqnarray}
where $\tau$ is the temporal resolution of the probe.

We see that both the mean value of the force and the correlations decay exponentially. However, the force fluctuations are highly non-Markovian, except for the limit $\nu = 0$, at which the force is trivially constant. The parameter $\Gamma$ defines not only the decay rate, but also the memory time of the non-Markovian process.

\subsection{Quantum harmonic-oscillator probe: superpositions}

Now consider a quantum probe made of a harmonic oscillator of frequency $\omega$ that is constrained to move along the horizontal axis as in Fig. 1.
The Hamiltonian of the harmonic oscillator probe is
\begin{equation}
H_P =  \omega \hat a^\dagger \hat a    \label{Hp}
\end{equation}
If the amplitude of the oscillations is much smaller than $L$, the length scale of the cat state,   the force acted upon the oscillator along the $x$ direction  is approximately constant and equal to Eq. (\ref{fxa}). This corresponds to an interaction Hamiltonian
\begin{eqnarray}
\hat{H}_I =  -f_0 \hat{\sigma}_3 \hat{x} = - \frac{f_0}{\sqrt{2m_0 \omega}} \hat{\sigma}_3 (\hat{a} + \hat{a}^{\dagger}),
\end{eqnarray}

 Thus, the total Hamiltonian of the two state system interacting with the oscillator probe is
\begin{eqnarray}
\hat{H} =  H_S + H_P + H_I = \n \hat{\sigma}_1 + \omega \hat{a}^{\dagger} \hat{a} + g \hat{\sigma}_3  (\hat{a} + \hat{a}^{\dagger}), \label{hjc}
\end{eqnarray}
where the system Hamiltonian $H_S$ is given by Eq. (\ref{hamqub}) specialized to $\chi = 0$,
i.e., it is equivalent to the Hamiltonian of a single-mode Jaynes-Cummings model with a coupling constant
\begin{eqnarray}
g = - \frac{f_0}{\sqrt{2m_0 \omega}}.
\end{eqnarray}

Often, the Jaynes-Cumming model is solved via the  Rotating Wave Approximation (RWA). However, the RWA is a good only near resonance and for small values of the coupling. This is not the regime of physical relevance here. The harmonic oscillator is supposed to be a probe, a pointer variable in the sense of quantum measurement theory. For this reason, the coupling must be the dominant term in the total Hamiltonian, as seen, for example, in the standard model of a von Neumann measurement  \cite{vNeu}.

This implies that we must consider the deep strong coupling regime \cite{CRLCR} of the Jaynes-Cummings model that corresponds to $g > \omega$. While the model has been recently shown to be integrable and the spectrum of the Hamiltonian has been computed \cite{braak}, it is still difficult to obtain closed  expressions for the dynamics  \cite{JCevol}. Here, we restrict to  an exactly solvable regime which corresponds to vanishing tunneling rate and to the limit of weak tunneling rate which can be accessed perturbatively. This will serve as a first approximation for the problem under consideration.

\subsubsection{Vanishing tunneling rate}

First, we  assume that $\n << \omega$, so that we can ignore the $\n \hat{\sigma}_1$ term in Eq. (\ref{hamqub}).  We call the resultant Hamiltonian $H_0$:
\begin{eqnarray}
\hat{H}_0 =  \omega \hat{a}^{\dagger} \hat{a} + g \hat{\sigma}_3  (\hat{a} + \hat{a}^{\dagger}), \label{hjc2}
\end{eqnarray}

This approximation defines the {\em adiabatic regime} of the Jaynes-Cummings model, which is fully solvable. The evolution operator is
\begin{eqnarray}
e^{-i\hat{H}_0t} = e^{i\frac{g^2}{\omega}t} \left( \begin{array}{cc} \hat{D}^{\dagger}(g/\omega) e^{-i\omega \hat{a}^{\dagger}\hat{a}t} \hat{D}(g/\omega) &0\\ 0&\hat{D}(g/\omega) e^{-i\omega \hat{a}^{\dagger}\hat{a}t} \hat{D}^{\dagger}(g/\omega) \end{array} \right) \label{evolut0}
\end{eqnarray}
where $\hat{D}(w) = e^{w\hat{a}^{\dagger} -w^*\hat{a}}$ stands for the displacement operator.

As initial condition at $t=0$, we assume that the oscillator is the vacuum state and the gravitational 2 state system is in a superposition state
 $c_+|+\rangle + c_-|-\rangle$.  Evolving with (\ref{evolut0}), we find the state of the system   at time $t$
\begin{eqnarray}
|\Psi(t)\rangle = e^{i\frac{g^2}{\omega^2}[\omega t - \sin (\omega t)] } \left(\begin{array}{c} c_+ |\zeta(t) \rangle  \\ c_- |-\zeta(t)\rangle\end{array} \right),
\end{eqnarray}
where the path
\begin{eqnarray}
\zeta(t) = -\frac{g}{\omega} (1 - e^{-i \omega t}),
\end{eqnarray}
describes an oscillation centered around $\zeta_0 = -\frac{g}{\omega} =  \frac{f_0}{\sqrt{2m_0\omega^3}} $. The center of the oscillation corresponds to  position  $x_0 = \frac{f_0}{m_0 \omega^2}$ and momentum $p_0 = 0$.


The reduced density matrix for the oscillator degrees of freedom is
\begin{eqnarray}
\hat{\rho}_{red}(t) &=& |c_+|^2 |\zeta(t)\rangle \langle \zeta(t)| + |c_-|^2 |-\zeta(t)\rangle \langle -\zeta(t)| \nonumber \\
&+& c_+^* c_2 |\zeta(t)\rangle \langle -\zeta(t)| + c_+^*c_-|-\zeta(t)\rangle\langle \zeta(t)| \label{reduced}
\end{eqnarray}
The off-diagonal terms are distinguishable only if $|\langle \zeta_0 |-\zeta_0\rangle|^2<<1$, or equivalently if $e^{-4|\zeta_0|^2}<< 1$. This implies that
\begin{eqnarray}
 \omega^3 << \left( \frac{f_0}{m_0}\right)^2m_0. \label{dist}
\end{eqnarray}
Eq. (\ref{dist}) is a necessary  condition for the oscillator   to act as a probe of the cat state. It demonstrates that the ultra-strong coupling limit of the Jaynes-Cummings model is the physically relevant regime.
For sufficiently large mass $m_0$, the oscillator may be a mesoscopic particle and its position at specific instant of time  measurable. This would imply a resolution $\sigma$ for the position measurement   smaller than the width $x_0$.

Hence, a position measurement at time $t$ has probability equal to $|c_+|^2$ of finding the oscillator in the neighborhood of the phase space point $\zeta(t)$ and probability equal to $|c_-|^2$ of  finding the oscillator in the neighborhood of the phase space point $-\zeta(t)$.

Of particular interest are the states $|\zeta_0, +\rangle = |\zeta_0\rangle \otimes |+\rangle$ and $|-\zeta_0, -\rangle = |-\zeta_0\rangle \otimes |-\rangle$ where $|\pm\zeta_0\rangle = \hat{D}(\zeta_0) |0\rangle$ is a  coherent state. We find that their time evolution involves only a time-dependent phase,
\begin{eqnarray}
e^{-i\hat{H}_0t}  |\pm \zeta_0, \pm \rangle = e^{i\frac{g^2}{\omega}t }|\pm \zeta_0, \pm \rangle,
\end{eqnarray}
 and thus they define stationary states of the system.

\subsubsection{Rabi oscillations}

A finite value of $\nu$ allows for transitions between the two gravitational quantum states, which induce transitions among the phase space paths of the oscillator. While the model is not exactly solvable, we can estimate the rate of such transitions using perturbation theory with respect to the tunneling rate $\nu$. In Appendix B, we show that to leading order in $\nu$, $e^{-i\hat{H}t} = e^{-i\hat{H}_0t} \hat{O}_t$, where
\begin{eqnarray}
\hat{O}_t  = \left( \begin{array}{cc} \cos \nu t & -i \sin \nu t \hat{D}(2\zeta_0) \\ - i\sin \nu t \hat{D}(-2\zeta_0) &\cos \nu t\end{array} \right). \label{otev}
\end{eqnarray}

As an estimate of the transition between the two gravitational quantum states, we compute the amplitude $\langle -\zeta_0, -|\hat{O}_t|\zeta_0, +\rangle$, between the stationary states $|\zeta_0, +\rangle$ and $|-\zeta_0, -\rangle$. We find
\begin{eqnarray}
\langle -\zeta_0, -|\hat{O}_t|\zeta_0, +\rangle = - i \sin \nu t,
\end{eqnarray}
and thus the associated probability
\begin{eqnarray}
p(t) = |\langle -\zeta_0, -|\hat{O}_t|\zeta_0, +\rangle|^2 = \sin^2 \nu t,
\end{eqnarray}
exhibits Rabi-type oscillations, with frequency $\nu$.

\section{Discussions}

The key findings of this work are enumerated in the Abstract.  We make some additional comments below.

 \subsubsection*{1. Comparison with alternative quantum theories.}
 At the Newtonian level, the gravitational field is completely slaved to the mass density.  Poisson's equation is the Newtonian limit of the Hamiltonian {\em constraint}  of General Relativity and not a dynamical equation.
 Thus, once we have a theory that determines the dynamics and  of mass density and its behavior under acts of measurement (in the standard quantum theory), we straightforwardly  obtain a description for the Newtonian gravitational field.

 In particular, we {\em need make no assumption that gravity is fundamentally quantum}, because the quantization of gravity refers to the true degrees of freedom and not to the gauge-dependent variables that appear in the constraints.   We only assume that there are no unknown physics  in the relation between mass density and gravitational forces in the Newtonian regime. An example otherwise is the Newton-Schrodinger equation---see, Refs.  \cite{AnHu14, BGDB, GiGr} for the relevant discussion.

 Thus, the foundational question that can be settled by experiments  invoking G2S systems is the following. {\em Does the gravitational force remain slaved to the mass density  even if the latter behaves quantum mechanically (i.e., when it fluctuates, or  subjected to quantum measurements)?}

The results derived here are based on standard quantum mechanics and (the nonrelativistic limit of) general relativity.  Both theories have stood the test of time in their respective domains of validity.  Results  presented here could serve as a useful benchmark for corresponding results from alternative quantum theories (AQT) to be compared and scrutinized.   We expect discrepancies, even  contradictions, appearing which should provide useful focal points for cross-examination of all AQTs in relation to the standard theories.

\subsubsection*{ 2.  Quantum Jumps of Spacetime Geometries.}    The cat state we consider here corresponds to the two minima of a potential, with a non-zero tunneling rate $\nu$ between them. A transition between two orthogonal qubit states  is standardly described as a quantum jump. The classic quantum jump experiments in atomic systems have shown that the duration of the jump is too small to be resolved \cite{NSD85}, so jumps are effectively instantaneous.  This implies the possibility of  instantaneous jumps between two spacetime geometries that correspond to different mass distributions. The "instantaneous" part is not an issue here, because Newtonian gravity admits action at a distance, and in the non-relativistic regime we cannot explore issues of causality.
However, {\em the idea that quantum jumps can occur in the spacetime description  because of the interaction of gravity with quantum matter} is a new phenomenon of  foundational value that is worthy of closer experimental exploration.

 \subsubsection*{3. From idealized models to realistic measurements.}The two measurement schemes we have considered correspond to {\em two major paradigms of quantum measurements}, namely, a continuous measurement and a von Neumann measurement. Our results also exemplify typical behaviors of probes.  We find that a classical probe records a  {\em non-Markovian fluctuating force}.  A quantum probe interacting with the G2S system  undergoes Rabi oscillations in the appropriate regime.   For follow-up work along the present line of inquiry,  state-of-the-art measurement schemes using atomic, molecular or light interferometric or electro- or opto-mechanical techniques would be very helpful.


\newpage

\begin{appendix}

\section{Probabilities and correlations functions for continuous measurement of the force}
We consider a continuous-time measurement of the  operator (\ref{fobs}) that corresponds to the gravitational force exerted upon a probe particle.

Let $\tau$ be the temporal resolution of the force measurements and let $T$ stand for the duration of the measurement. We decompose the interval $[0, T]$ into $N = T/\tau$ subintervals, or equivalently $N+1$ time steps, $i = 0, 1, 2, \ldots, N$. At each time step, there are two possible outcomes to the measurement, corresponding either to $a = +$ or to $a = -$ in Eq. (\ref{fxa}).  Without loss of generality we assume the initial state to be $|+\rangle$, i.e., $c_+ = 1, c_- = 0$.

A recorded series of values is represented by $n+1$ integers $k_i$ that satisfy  $\sum_{i=0}^{n} k_i = N+1$,  so that the first $k_0$ entries of the series are $+$, the following $k_1$ entries are $-$, and so on until the final $k_n$ entries that are $(-1)^n$. The integer $n \leq N$ stands for the  number of jumps from $+$ to $-$ or the reverse.

The probability associated with a sequence of $N$ records is
\begin{eqnarray}
  P(a_1, t_1; a_2, t_2; \ldots; a_N t_n) = Tr \left[\hat{P}_{a_Nt_N}\ldots \hat{P}_{a_2t_2} \hat{P}_{a_1t_1}\hat{\rho}_0 \hat{P}_{a_1t_1} \hat{P}_{a_2t_2} \ldots \hat{P}_{a_nt_n}\right] \label{ntimepr2b},
\end{eqnarray}
where $a_i = \pm$.

Taking times $t_i = i \tau$, $i = 1, \ldots N$ and representing the sequence of records by the $n$ integers $k_i$, we obtain
\begin{eqnarray}
P(k_1, k_2, \ldots, k_n) = |\langle +| \hat{U}_{\tau}|+\rangle|^{2(N-n)} |\langle -| \hat{U}_{\tau}|+\rangle|^{2n}, \label{ntimepr3a}
\end{eqnarray}
where we have used the fact that  $ \langle +| \hat{U}_{\tau}|+\rangle =   \langle -| \hat{U}_{\tau}|-\rangle $ in a symmetric confining potential. Remarkably, the probability (\ref{ntimepr3a}) {\em depends only on the number $n$ of jumps}.

Since
$ |\langle +| \hat{U}_{\tau}|+\rangle|^2 = \cos^2\frac{\nu\tau}{2} $ and
$  |\langle -| \hat{U}_{\tau}|+\rangle|^2 = \sin^2\frac{\nu\tau}{2}$, Eq. (\ref{ntimepr3a}) becomes
 \begin{eqnarray}
 P(k_1, k_2, \ldots, k_n) = \left(\cos\frac{\nu\tau}{2}\right)^{2(N-n)}  \left(\sin\frac{\nu\tau}{2}\right)^{2n}.\label{ntimepr3b}
 \end{eqnarray}
In the regime  $\nu \tau << 1$,
   \begin{eqnarray}
 P(k_1, k_2, \ldots, k_n) = \lambda^n e^{-\lambda(N-n)}, \label{ntimepr3c}
 \end{eqnarray}
 where
 \begin{eqnarray}
 \lambda = \frac{\nu^2 \tau^2}{4}.  \label{lambd}
 \end{eqnarray}
 In order to find the correlation functions associated with the probabilities (\ref{ntimepr3c}), we compute the conditional probability $P(a_2, i_2| a_1,i_1)$ that for $a_2$ at time step $i_2$  given a measurement outcome $a_1$ at time step $i_1$. This equals the sum of the probabilities (\ref{ntimepr3c}) over all sequences $a_i$ over time steps $i$, such that $i_1 \leq i \leq i_2$, such that $a_{i_1} = a_1$ and $a_{i_2} = a_2$.

To calculate this we  first note that the probabilities (\ref{ntimepr3c})  do not depend on any initial moment of time, so the conditional probability is a function of $m = i_2-i_1$. Thus, we need to compute a function $g(a_2,a_1;m) = P(a_2, i_2| a_1,i_1)$.

 First, consider $a_1 = a_2= +$. Only sequences with an even number $n = 2l$ of jumps contribute. The number of different sequences of length $m+1$ that start from $+$ and contain $n$ jumps is $\binom{m}{n}  $. Hence,
 \begin{eqnarray}
 g(+,+;m) = \left(\cos\frac{\nu\tau}{2}\right)^{2m} \sum_{l=0}^{[m/2]} \binom{m}{2l}    \left(\frac{1}{2}\sin \nu\tau\right)^{4l} \label{g++}
 \end{eqnarray}
 Similarly, we find
 \begin{eqnarray}
 g(-,+;m) = \left(\cos\frac{\nu\tau}{2}\right)^{2m}  \sum_{l=0}^{[(m-1)/2]}\binom{m}{2l+1}      \left(\frac{1}{2}\sin \nu\tau\right)^{4l+2} \label{g--}
 \end{eqnarray}

We employ the binomial identity  $(1+x)^m = \sum_{l=0}^m  \binom{m}{l}   x^l$, to obtain
 \begin{eqnarray}
  g(+,+;m) = \frac{1}{2} \left(\cos\frac{\nu\tau}{2}\right)^{2m} \left[ (1+\frac{1}{4}\sin^2 \nu\tau)^m + (1 - \frac{1}{4}\sin^2 \nu\tau)^m\right]\\
   g(-,+;m) = \frac{1}{2} \left(\cos\frac{\nu\tau}{2}\right)^{2m} \left[ (1+\frac{1}{4}\sin^2 \nu\tau)^m -(1 - \frac{1}{4} \sin^2 \nu\tau)^m\right]
 \end{eqnarray}
 By symmetry, $g(-,-;m) = g(+,+;m)$ and   $g(+,-;m) =   g(-,+;m) $.

We  assume an initial state $|+\rangle$. The expectation value of the force at the time-step $m$ is
\begin{eqnarray}
\langle F(m)\rangle = - f_0 [  g(+,+;m) -   g(-,+;m)] = - f_0 \left[\cos^2\frac{\nu\tau}{2} (1 - \frac{1}{4}\sin^2 \nu\tau)\right]^{m}
\end{eqnarray}
The correlation function at time steps $m_1$ and $m_2 > m_1$ is
\begin{eqnarray}
\langle F(m_2) F(m_1)\rangle = f_0^2 \left[g(+,+;m_2-m_1) - g(-,+;m_2-m_1)  \right]\nonumber \\
\times
\left[ g(+,+;m_1) + g(-,+;m_1) \right]  \nonumber \\
= f_0^2 \left(\cos^2\frac{\nu\tau}{2}\right)^{m_2}(1 - \frac{1}{4}\sin^2 \nu\tau)^{m_2 - m_1} (1+\frac{1}{4}\sin^2 \nu\tau)^{m_1}
\end{eqnarray}
In the regime $\nu \tau << 1$,
\begin{eqnarray}
\langle F(m)\rangle = - f_0 e^{-2\lambda m} \\
\langle F(m_2) F(m_1)\rangle = f_0^2 e^{-2\lambda (m_2-m_1)},
\end{eqnarray}
where $\lambda$ is given by Eq. (\ref{lambd}).

The continuous-time limit follows by setting $m = t /\tau$ and defining the decay constant $\Gamma = 2 \lambda/\tau =  \frac{\nu^2\tau}{2}$. Then, Eqs. (\ref{corrcon1}, \ref{corrcon2}) follow.

\newpage

\section{Perturbative propagator for the Jaynes-Cumming model}

We compute the evolution operator $e^{-i\hat{H}t}$, where $\hat{H}$ is given by Eq. (\ref{hjc}). We write $\hat{H} = \hat{H}_0 + \hat{V}$, where  $\hat{H}_0$ is given by Eq. (\ref{hjc2}) and
 $\hat{V} = \nu \sigma_1$. We work in the interaction representation. We define the unitary operator
\begin{eqnarray}
\hat{O}_t = e^{i\hat{H}_0t} e^{-i\hat{H}t},
\end{eqnarray}
which  is a solution to the equation
\begin{eqnarray}
i\frac{\partial}{\partial t}\hat{O}_t = \hat{V}(t) \hat{O}_t, \label{vtnu}
\end{eqnarray}
with initial condition $\hat{O}_0 = \hat{1}$. In Eq. (\ref{vtnu})
\begin{eqnarray}
\hat{V}(t) = \nu e^{i\hat{H}_0t} \hat{\sigma}_1 e^{-i\hat{H}_0t} = \left( \begin{array}{cc} 0 & \hat{S}_t \\ \hat{S}^{\dagger}_t &0\end{array} \right).
\end{eqnarray}
in terms of
\begin{eqnarray}
\hat{S}_t  = \hat{D}(\zeta_0) e^{i\hat{H}_0t} \hat{D}(-2\zeta_0) e^{-i\hat{H}_0t}\hat{D}(\zeta_0)
\end{eqnarray}
where $D(z)$ is the standard displacement operator.

To leading order in perturbation theory, the solution $\hat{O}_t$ can be expressed as
\begin{eqnarray}
\hat{O}_t = e^{-i \int_0^t ds \hat{V}(s)}, \label{ot2}
\end{eqnarray}
i.e., the time-ordered exponential that solves Eq. (\ref{vtnu}) is substituted by an ordinary exponential. We find
\begin{eqnarray}
\int_0^t ds \hat{S}(s) = \hat{D}(\zeta_0) \left(\int_0^tds e^{-2\zeta_0 (\hat{a}^{\dagger}e^{i\omega t} - \hat{a} e^{-i \omega t}) }\right) \hat{D}(\zeta_0).
\end{eqnarray}
For $\omega t >> 1$, the time integration over a periodic function is negligible, so  $\int_0^tds e^{-2\zeta_0 (\hat{a}^{\dagger}e^{i\omega t} - \hat{a} e^{-i \omega t}) } \simeq t$. Hence, $\int_0^t ds \hat{S}(s)= \hat{D}(2\zeta_0)$, and
\begin{eqnarray}
\int_0^t ds \hat{V}(s) = \nu t  \left( \begin{array}{cc} 0 & \hat{D}(2\zeta_0) \\ \hat{D}(-2\zeta_0) &0\end{array} \right).
\end{eqnarray}
By Eq. (\ref{ot2}), we obtain Eq. (\ref{otev}) for $\hat{O}_t$.

\end{appendix}

\end{document}